\input mtexsis
\input epsf
\paper
\singlespaced
\widenspacing
\twelvepoint
\Eurostyletrue
\thicksize=0pt
\sectionminspace=0.1\vsize

\def\nc{{N_c}}
\def\yo1{{F_\pi^2}}
\def\llrb{{\rightarrow}}
\def\llra{{\Rightarrow}}
\def\mapright#1{{\smash{\mathop{\llra}\limits_{#1}}}}
\def\maprightb#1{{\smash{\mathop{\llrb}\limits_{#1}}}}

\def\sss{\scriptscriptstyle}


\referencelist
\reference{exp1}
D.R.~Thompson {\it et. al.}  [E852 Collaboration],
\journal Phys. Rev. Lett.;79, 1630 (1997), {{\tt hep-ex/9705011}};
A.~Abele {\it et. al.}  [Crystal Barrel Collaboration],
\journal Phys. Lett. B;423, 175 (1998)
\endreference
\reference{exp2}
G.S.~Adams {\it et. al.}  [E852 Collaboration],
\journal Phys. Rev. Lett.;81, 5760 (1998)
\endreference
\reference{thooft} G.~'t Hooft,
\journal Nucl. Phys. B;72,461 (1974);
E.~Witten, \journal Nucl. Phys. B;156,269 (1979)
\endreference
\reference{Cohen}
T.D.~Cohen,
\journal Phys. Lett. B;427,348 (1998), {{\tt hep-ph/9801316}}
\endreference
\reference{lattice} P.~Lacock {\it et. al.} [UKQCD Collaboration],
\journal Phys. Rev. D;54,6997 (1996), \hfill\break {{\tt hep-lat/9605025}}
\endreference
\reference{*latticea} 
C.~Bernard {\it et. al.}  [MILC Collaboration],
\journal Phys. Rev. D;56, 7039 (1997), \hfill\break {{\tt hep-lat/9707008}}
\endreference
\reference{*latticeb} 
P.~Lacock and K.~Schilling  [SESAM collaboration],
\journal Nucl. Phys. Proc. Suppl.;73, 261 (1999), {{\tt hep-lat/9809022}}
\endreference
\reference{Swanson}
T.~Barnes, F.E.~Close and E.S.~Swanson,
\journal Phys. Rev. D;52, 5242 (1995), \hfill\break {{\tt hep-ph/9501405}}
\endreference
\reference{Michael}
C.~Michael, {{\tt hep-ph/0101287}}
\endreference
\reference{Golterman}
M.~Golterman and S.~Peris, 
\journal JHEP;0101,028 (2001), {{\tt hep-ph/0101098}}
\endreference
\reference{Beanelast}
S.R.~Beane, {{\tt hep-ph/0106022}}
\endreference
\reference{Bramon}
A.~Bramon, E.~Etim and M.~Greco,
\journal Phys. Lett. B;41,609 (1972);
M.~Greco,
\journal Nucl. Phys. B;63,398 (1973);
J.~J.~Sakurai,
\journal Phys. Lett. B;46,207 (1973);
E.~Etim, M.~Greco and Y.~Srivastava,
\journal Lett. Nuovo Cim.;16,65 (1976);
B.~V.~Geshkenbein,
\journal Sov. J. Nucl. Phys;49,705 (1989)
\endreference
\reference{thooft2} 
G.~'t Hooft,
\journal Nucl. Phys. B;75, 461 (1974)
\endreference
\reference{cole} S.~Coleman and E.~Witten,
\journal Phys. Rev. Lett.;45,100 (1980)
\endreference
\reference{Shifman}
M.A.~Shifman, A.I.~Vainshtein and V.I.~Zakharov,
\journal Nucl. Phys. B;147,385 (1979)
\endreference
\reference{pdg}
D.E.~Groom {\it et. al.}  
\journal Eur. Phys. J. C;15,1 (2000)
\endreference
\reference{Beane2}
S.R.~Beane,
\journal Phys. Rev. D;61,116005 (2000), {{\tt hep-ph/9910525}}
\endreference
\reference{Stern}
M.~Davier, L.~Girlanda, A.~Hocker and J.~Stern,
\journal Phys. Rev. D;58,096014 (1998), {{\tt hep-ph/9802447}}
\endreference
\reference{Indu}
F.J.~Yndurain,
\journal Phys. Rept.;320,287 (1999), {{\tt hep-ph/9903457}}
\endreference
\reference{Teper}
M.J.~Teper, {{\tt hep-lat/9711011}}
\endreference
\reference{GF11}
F.~Butler, H.~Chen, J.~Sexton, A.~Vaccarino and D.~Weingarten,
\journal Nucl. Phys. B;430, 179 (1994), {{\tt hep-lat/9405003}}
\endreference
\reference{UKQCD}
R.~Kenway  [UKQCD Collaboration],
\journal Nucl. Phys. Proc. Suppl.;53, 206 (1997), {{\tt hep-lat/9608034}}
\endreference
\reference{*UKQCDa}
H.P.~Shanahan {\it et. al.}  [UKQCD Collaboration],
\journal Phys. Rev. D;55, 1548 (1997), {{\tt hep-lat/9608063}}
\endreference
\reference{Govaerts}
J.~Govaerts, F.~de Viron, D.~Gusbin and J.~Weyers,
\journal Nucl. Phys. B;248, 1 (1984)
\endreference
\reference{*Govaertsa}
I.I.~Balitsky, D.~Diakonov and A.V.~Yung,
\journal Z. Phys. C;33, 265 (1986)
\endreference
\reference{*Govaertsb}
J.I.~Latorre, P.~Pascual and S.~Narison,
\journal Z. Phys. C;34, 347 (1987)
\endreference
\endreferencelist

\titlepage
\obeylines
\hskip4.8in{NT@UW-01-019}\unobeylines
\vskip0.5in
\title
Quark-Hadron Duality for Hybrid Mesons at Large-$\nc$ 
\endtitle
\author
Silas R.~Beane

Department of Physics, University of Washington
Seattle, WA 98195-1560
\vskip0.1in
\center{{\it sbeane@phys.washington.edu}}\endcenter
\endauthor

\abstract
\singlespaced
\widenspacing

We investigate implications of quark-hadron duality for hybrid mesons in the
large-$\nc$ limit. A simple formalism is developed which implements duality for
QCD two-point functions of currents of quark bilinears, with any number of
gluons. We argue that the large-$\nc$ meson masses share a common parameter,
which is related to the QCD string tension. This parameter is fixed from
correlators of conserved vector and axial-vector currents, and using lattice
QCD determinations of the string tension. Our results predict towers of hybrid
mesons which, within expected $1/\nc$ corrections, naturally accommodate the
$1^{-+}$ experimental hybrid candidates.

\endabstract 
\vskip0.5in
\endtitlepage
\vfill\eject                                     
\superrefsfalse
\singlespaced
\widenspacing

\vskip0.1in
\noindent {\twelvepoint{\bf 1.\quad Introduction}}
\vskip0.1in

\noindent The possible experimental observation of mesons with quantum numbers
that cannot arise from a simple quark-antiquark state, has generated a great
deal of interest\ref{exp1}\ref{exp2}.  In effect, search for these hybrid meson
states is a primary mission of the proposed Hall D at Jefferson Laboratory.
There are currently two experimental $J^{PC}=1^{-+}$ hybrid candidates, at
$\sim 1.4~{\rm GeV}$\ref{exp1} and $\sim 1.6~{\rm GeV}$\ref{exp2}.  One reason
for excitement among theorists is the possibility that these states offer a
direct experimental probe of the gluonic nature of QCD.  Of course, {\it a
priori} there is no reason why these meson states should not couple strongly to
four-quark currents, rather than to currents involving gluons. In this respect,
the large-$\nc$ limit comes into its own as an extremely useful
tool\ref{thooft}\ref{Cohen}.

In the large-$\nc$ limit, mesons are narrow and couple to QCD
currents with one quark, one antiquark and any number, $n_g$, of
gluons\ref{thooft}\ref{Cohen}. We define a hybrid to be a meson whose
space-time quantum numbers require $n_g\geq 1$. As with conventional
mesons\ref{thooft}, the QCD correlators of currents which couple to hybrid
mesons can be expressed as sums of meson tree graphs, which must be infinite in
order to be dual to perturbative QCD at large momentum transfer\ref{Cohen}.
By contrast, quenched lattice QCD calculations\ref{lattice} suggest that the lightest 
hybrid has $1^{-+}$ and a mass $\sim 2~{\rm GeV}$. 
(A survey of results from hadronic models is available in \Ref{Swanson}.)
The coupling of the hybrid mesons to currents with four-quark states is
suppressed in the quenched lattice calculations. One might expect that
the discrepancy between the lattice and experiment is due to this
suppression\ref{Michael}\ref{lattice}. The natural forum to test this 
hypothesis is the large-$\nc$ limit, as the coupling of hybrid mesons to 
four-quark currents is subleading in the $1/\nc$ expansion\ref{thooft}\ref{Cohen}.

In this paper we investigate duality in the large-$\nc$ limit for QCD
correlators with hybrid quantum numbers. We develop a simple formalism which
implements duality for arbitrary two-point functions of currents of quark bilinears.
The Wilson coefficients in the operator product expansion (OPE) of correlators
with $n_g=1$ also appear in the OPE of correlators with $n_g=0$.  We also argue
that the large-$\nc$ meson masses depend on a universal parameter, $\Lambda$,
which is related to the QCD string tension.  Recent work has investigated
quark-meson duality in the large-$\nc$ limit for conserved vector and
axial-vector QCD currents\ref{Golterman}\ref{Beanelast}~(see also
\Ref{Bramon}). We use this phenomenology of ordinary mesons as input into our
investigation of hybrids.

In Section 2, we develop a simple formalism for investigating quark-meson
duality in the large-$\nc$ limit. In Section 3 we apply this formalism to
conserved vector and axial-vector QCD currents and their well-known
phenomenology. This section is a necessary review of work that has appeared
elsewhere\ref{Beanelast}. We discuss the issue of setting the scale at which
to evaluate the QCD coupling constant in Section 4. We also introduce a
universality conjecture in this section. In Section 5 we investigate vector and
axial-vector currents which couple to hybrid mesons. We discuss our results 
in Section 6.

\vskip0.1in
\noindent {\twelvepoint{\bf 2.\quad The General Statement of Duality}}
\vskip0.1in

\noindent The basic object we will consider in this paper is the two-point 
function, 

$$ \Pi(Q^2)=i\int {d^4x}{e^{iqx}} \bra{0}T\lbrack{\rm J}^\dagger (x){\rm
J}(0)\rbrack\ket{0}\, , \EQN symbdef$$ 
where $J(x)$ is a color-singlet QCD current of the form 
${\bar q}\,\Gamma (n_g)\,q$, where 
$\Gamma (n_g)$ is an arbitrary Dirac-Lorentz structure coupled to $n_g$ gluons, and
${Q^2}=-{q^2}$. Here for illustration we have assumed a scalar current with
respect to all quantum numbers.  The general form of the spectral function of
mass-dimension $2m$ in the large-$\nc$ limit is

$$ {1\over\pi} Im \,\Pi(t)= 2\sum_{n=0}^N {F^2}(n) M^{2m}(n) \delta (t-M^2 (n)) ,
\EQN spectfun$$ 
where $F(n)$ is the decay constant of the $n$th meson, $M(n)$ is its mass,
and $N$ regulates the number of mesons, which is strictly speaking, infinite. This
spectral function expresses the statement that in the large-$\nc$ limit, 
$\Pi(Q^2)$ is given by an infinite sum of zero-width mesons.
We can now construct $\Pi(Q^2)$ using dispersion theory.
At large space-like momenta we obtain the general duality relation

$$\eqalign{ 
&2(-1)^m Q^{2m}\, {\sum _{n=0}^N}
{{{F^2}(n)}\over{{Q^2}+{M^2}(n)}}\, +\,
\sum_{k=0}^{m} a(\mu ;N)_k Q^{2k}\cr
&\qquad\qquad\,\mapright{{{Q^2\,\rightarrow\,\infty}}}\,
{\cal C} (\mu )Q^{2m}\;\log{{Q^2}\over{\mu^2}}\,+\,
{\sum_{\ell=0}^{\infty}} 
{{{\vev{{\cal O}(Q^2,\mu )}}
\over{Q^{2\ell}}}}^{\sss d=2(\ell +m)}+\ldots}
\EQN full$$ 
where ${\cal C}$ is an analytic function of $\alpha_s$ which is computed in QCD
perturbation theory and  $\vev{{\cal O}(Q^2,\mu )}$ are Wilson coefficients of
mass-dimension $d=2(\ell +m)$. The $a(N;\mu)_k$ are hadronic counterterms which
render the hadronic sums over states independent of $N$, and $\mu$ is a
renormalization scale defined in dimensional regularization with
$\overline{MS}$. The dots represent higher order perturbative contributions.
(For a diagrammatic statement of this duality relation, see \Fig{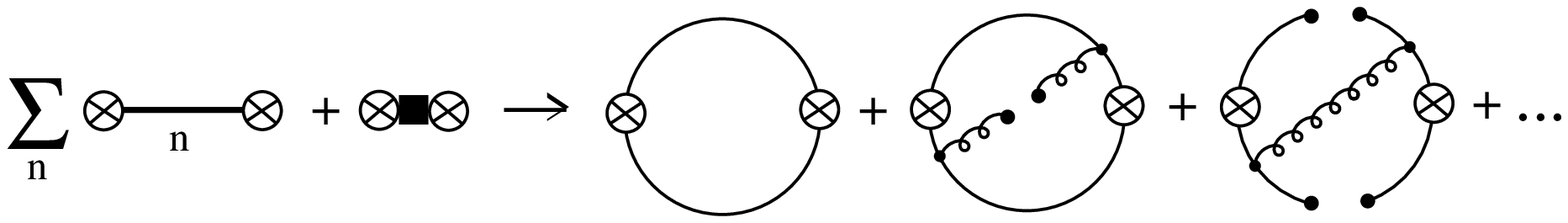} and
\Fig{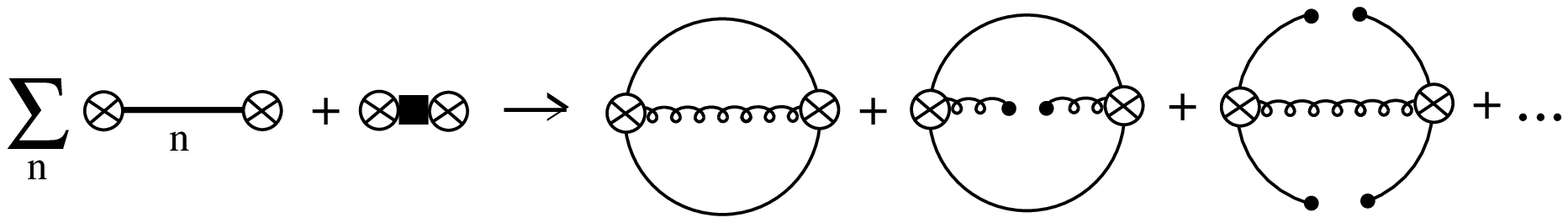}.)  \Eq{full} implies 

$$ {{{F^2}(n)}/{{M^2}(n)}}\quad\maprightb{{n\,\rightarrow\,\infty}}
\quad{n^{-1}}.  \EQN limt$$ 
In this paper we will assume that the squared masses are linear in $n$. 
This is the case in two-dimensional QCD in the large-$\nc$ limit\ref{thooft2}. 
And, too, linearity in $n$  is expected if large-$\nc$ QCD is in some sense 
equivalent to a string theory. There is a caveat to this assumption relating to
chiral symmetry breaking, which will be addressed below.  It follows from \Eq{limt} that
$F^2(n)=F^2 +{\rm O}(n^{-\epsilon})$ where $F$ is a hadronic parameter and
$\epsilon$ is a positive number. We will keep the leading correction to $F^2$.
We choose the parameterization

$$\EQNalign{&M^2(n)=M^2 +\Lambda^2 n  \EQN cacauno;a \cr
&F^2(n)=F^2 + {{E^{2+2\epsilon}}\over{M^{2\epsilon}(n)}}.
\EQN cacauno;b}
$$
Hence, {\it a priori}, each QCD two-point function is parameterized by four
hadronic parameters: $M$, $\Lambda$, $F$ and $E$.  It is straightforward to
convince oneself that in order that the leading Wilson coefficient ($d=2m$) not vanish,
we must have $\epsilon =m$.  On the hadronic side of \Eq{full} we then have

$$\EQNalign{&2(-1)^m \, Q^{2m}\, {\sum _{n=0}^N}
{{{F^2}}\over{{Q^2}+{M^2}(n)}}\, +\,
2{\sum _{n=0}^N}
{{{E^{2+2m}}}\over{{Q^2}+{M^2}(n)}}\, +\,
\sum_{k=0}^{m} {\bar a}(N;\mu)_k Q^{2k}=\cr
&-2( (-1)^{m} \, Q^{2m}\, {{F^2}\over{\Lambda^2}} \,+\,
{{E^{2+2m}}\over{\Lambda^2}})
\; (\Psi ({{Q^2 +M^2}\over{\Lambda^2}})-\Psi (N))
\, +\, \sum_{k=0}^{m} {\bar a}(N;\mu)_k Q^{2k}
\EQN psifunctions}$$ 
where $\Psi$ is Euler's function, defined as the logarithmic derivative of
the Gamma function. The counterterms here are barred, indicating that they
differ from those in \Eq{full}. From \Eq{full} we immediately find

$$\EQNalign{& {\cal C}(\mu )=(-1)^{m+1}{2{F^2}\over{\Lambda^2}}\, ,
\EQN match1;a \cr
&{\vev{{\cal O}(Q^2,\mu )}^{\sss d=2m}}={\cal W}(\mu )
\;\log{{Q^2}\over{\mu^2}}\, ,\qquad {\cal W}(\mu )=
-{2{E^{2+2m}}\over{\Lambda^2}} .
\EQN match1;b \cr}
$$
Using \Eq{full} and \Eq{psifunctions}, we can determine the counterterms, 
${\bar a}(N;\mu)_k$, and match to the OPE separately for each $m$. Working to 
leading order in $\alpha_s$ and keeping Wilson coefficients up to dimension six, we find 

$$\EQNalign{& 
{\bar a}(N;\mu)_0\,=\, {\cal C}(\mu )\,\log{{N{\Lambda^2}}\over{\mu^2}}\, ,
\EQN matchm0;a \cr
& \vev{{\cal O}(\mu )}^{\sss d=2}\,=\, {\cal C}(\mu )({M^2}-{1\over 2}{\Lambda^2})\, ,
\EQN matchm0;b \cr
& \vev{{\cal O}(\mu )}^{\sss d=4}\,=\, -{1\over 2}{\,}{\cal C}(\mu )
({M^4}-{M^2}{\Lambda^2}+{1\over 6}{\Lambda^4})\, ,
\EQN matchm0;c \cr
& \vev{{\cal O}(\mu )}^{\sss d=6}\,=\, {1\over 3}{\,}{\cal C}(\mu )
({M^2}-{1\over 2}{\Lambda^2})({M^2}-{\Lambda^2}){M^2}\, ,
\EQN matchm0;d \cr}
$$
for $m=0$, and

$$\EQNalign{& 
{\bar a}(N;\mu)_0\,=\, {\cal W}(\mu )\,\log{{N{\Lambda^2}}\over{\mu^2}}
+{1\over 2}{\,}{\cal C}(\mu )
({M^4}-{M^2}{\Lambda^2}+{1\over 6}{\Lambda^4})\, ,
\EQN matchm2;a \cr
&{\bar a}(N;\mu)_1\,=\, -{\cal C}(\mu )\,({M^2}-{1\over 2}{\Lambda^2})\, ,
\EQN matchm2;b \cr
&{\bar a}(N;\mu)_2\,=\, {\cal C}(\mu )\,\log{{N{\Lambda^2}}\over{\mu^2}}\, ,
\EQN matchm2;c \cr
& \vev{{\cal O}(\mu )}^{\sss d=6}\,=\, 
({1\over 3}{\,}{\cal C}(\mu )({M^2}-{\Lambda^2}){M^2}+{\cal W}(\mu ))({M^2}-{1\over 2}{\Lambda^2})\, ,
\EQN matchm2;d \cr}
$$
for $m=2$. We will consider these two cases in this paper.

One may wonder about the effects of higher-order contributions to $F^2(n)$.  An
${\rm O}(n^{-m-2})$ correction, parameterized by ${{D^{4+2m}}/{M^{2m+2}}(n)}$,
generates a $\log Q^2$ correction to the Wilson coefficient with
$d=2(1+m)$. This correction is higher order in $\alpha_s$ than the order to
which we are working.

\vskip0.1in
\noindent {\twelvepoint{\bf 3.\quad Conserved Vector and Axial-Vector Currents}}
\vskip0.1in

\noindent We consider QCD with two massless flavors.
This theory has an $SU(2)\times SU(2)$ chiral symmetry.  In the
large-$\nc$ limit, assuming confinement, this symmetry breaks
spontaneously to the $SU(2)_V$ isospin subgroup\ref{cole}. (Strictly
speaking, large-$\nc$ QCD has a $U(2)\times U(2)$ chiral symmetry.
In this paper the additional Goldstone mode and its associated $U(1)$
charge are ignored.) Consider the conserved QCD vector and axial-vector currents

\offparens
$$\EQNalign{ 
&{{\rm V}_{\mu}^a}=
{\bar {q}}{\gamma_\mu}{\textstyle{1\over 2}}{\tau^a}{q}\, ,
\EQN aandvdefinqcd;a \cr
&{{\rm A}_{\mu}^a}=
{\bar {q}}{\gamma_\mu}{\gamma_5}{\textstyle{1\over 2}}{\tau^a}{q}.
\EQN aandvdefinqcd;b \cr}
$$ 
These currents fill out a six-dimensional chiral multiplet;
they transform as $(\bf{3},\bf{1})\oplus (\bf{1},\bf{3})$ with respect to
$SU({2})\times SU({2})$.  We are interested in the correlators

$$\EQNalign{
&2i\int {d^4x}{e^{iqx}}
\bra{0}T\lbrack{{\rm V}_{a}^\mu}(x){{\rm V}_{b}^\nu}(0)\rbrack\ket{0}= {\delta_{ab}}
({q^\mu}{q^\nu}-{g^{\mu\nu}}{q^2}){\Pi_{\sss V}}({Q^2})\, ,
\EQN correl1;a \cr
&2i\int {d^4x}{e^{iqx}}
\bra{0}T\lbrack{{\rm A}_{a}^\mu}(x){{\rm A}_{b}^\nu}(0)\rbrack\ket{0}= {\delta_{ab}}
({q^\mu}{q^\nu}-{g^{\mu\nu}}{q^2}){\Pi_{\sss A}}({Q^2})\, . 
\EQN correl1;a \cr}
$$ 
Since $\Pi_{\sss V,A}({Q^2})$ are dimensionless, we are here
interested in the case $m=0$.  

Chiral symmetry constrains duality. Since chiral symmetry is not broken in
perturbation theory, $\Pi_{\sss V}(Q^2 )=\Pi_{\sss A}(Q^2 )$ to all orders in
$\alpha_s$.  Moreover, because the pion couples to the axial-vector current,
consistency with chiral symmetry requires that the pion, together with its
chiral partners, be left out of the tower of states\ref{Golterman}\ref{Beanelast}. The minimal
realistic ansatz for the spectral functions consistent with chiral symmetry is

$$\EQNalign{ 
&{1\over\pi} Im \,\Pi_{\sss V}(t)= 2{F_{\rho}^2} \delta (t-M_{\rho}^2 ) +
2\sum_{n=0}^{N_{\sss V}} {F_{\sss V}^2}(n) \delta (t-M_{\sss V}^2 (n)) \, ,
\EQN spectfunvand;a \cr
&{1\over\pi} Im \,\Pi_{\sss A}(t)= 2{F_{\pi}^2}\delta (t) +
2{F_{{a_1}}^2} \delta (t-M_{{a_1}}^2 ) +
2\sum_{n=0}^{N_{\sss A}} {F_{\sss A}^2}(n) \delta (t-M_{\sss A}^2 (n)) .
\EQN spectfunvand;b \cr} $$ 
Here we have extracted the lowest-lying vector and axial-vector mesons, $\rho$
and ${a_1}$, respectively. These states, together with the pion, satisfy
spectral function sum rules. This is equivalent to the statement that $\pi$,
$\rho$ and $a_1$, together with an isoscalar $S_0$, fill out a reducible
$(\bf{1},\bf{3})\oplus(\bf{3},\bf{1})\oplus(\bf{2},\bf{2})$ representation of
the chiral group\ref{Beanelast}. The parameters are then related by a single
mixing angle, $\phi$, via ${F_\pi}={F_\rho}\sin\phi$,
${F_{a_{1}}}={F_\pi}\cot\phi$ and ${M_\rho}={M_{a_{1}}}\cos\phi$.  One can
further show that each vector meson in the infinite sum must be paired with a
{\it degenerate} axial-vector chiral partner so that pair-by-pair they fill out
{\it irreducible} $(\bf{1},\bf{3})\oplus(\bf{3},\bf{1})$ representations of the
chiral group.  It follows that ${F_{\sss V}}={F_{\sss A}}\equiv F$,
${\Lambda_{\sss V}}={\Lambda_{\sss A}}\equiv \Lambda$ and ${M_{\sss
V}}={M_{\sss A}}\equiv M$~\ref{Beanelast}. It is intriguing that only the
states outside of the tower exhibit mass splittings.

\figure{vanda.eps}
\epsfxsize 6.5in
\centerline{\epsfbox{vanda.eps}}
\caption{\tenpoint{The statement of duality for the correlators of conserved vector and
axial-vector currents. The crossed circles represent an insertion of a
current. The thick line is a meson propagator and the filled-in square represents a
hadronic counterterm. The thin and wavy lines are quark and gluon propagators,
respectively. The small filled-in circles denote a perturbative interaction
and the large filled-in circles denote a condensate. We have not included permutations.}}
\endcaption
\endfigure

Reading directly from \Eq{match1;a} and \Eq{matchm0} we then find the system of equations

\autoparens
$$\EQNalign{ 
&{\cal C}_{\sss {V,A}}=-{\nc\over{12{\pi^2}}}\; ,
\EQN opejunk;a\cr
&{\vev{{\cal O}}^{\sss d=2}_{\sss {V,A}}}=0=
2{F_{\pi}^2}\csc^2\phi+{\cal C}_{\sss {V,A}}({M^2}-{1\over 2}{\Lambda^2})\; ,
\EQN opejunk;b\cr
&{\vev{{\cal O}}^{\sss d=4}_{\sss {V,A}}}={{\alpha_s}\over{12\pi}}\vev{G_{\mu\nu}G_{\mu\nu}}=
-2{F_{\pi}^2}{M_{\rho}^2}\csc^2\phi-{1\over 2}{\cal C}_{\sss {V,A}}
({M^4}-{M^2}{\Lambda^2}+{1\over 6}{\Lambda^4})\; ,
\EQN opejunk;c\cr
&{\vev{{\cal O}}^{\sss d=6}_{\sss {V}}}=
-{28\over 9}{\pi{\alpha_s}}\vev{{\bar q} q}^2=
2{F_{\pi}^2}{M_{\rho}^4}\csc^2\phi
+{1\over 3}{\cal C}_{\sss {V,A}}
({M^2}-{1\over 2}{\Lambda^2})({M^2}-{\Lambda^2}){M^2}\; ,
\EQN opejunk;d\cr
&{\vev{{\cal O}}^{\sss d=6}_{\sss {A}}}=
{44\over 9}{\pi{\alpha_s}}\vev{{\bar q} q}^2=
2{F_{\pi}^2}{M_{\rho}^4}{{\csc^2\phi}\over{\cos^2\phi}}
+{1\over 3}{\cal C}_{\sss {V,A}}
({M^2}-{1\over 2}{\Lambda^2})({M^2}-{\Lambda^2}){M^2}.
\EQN opejunk;e\cr}
$$ The Wilson coefficients and ${\cal C}_{\sss {V,A}}$ have been computed from
the diagrams in \Fig{vanda.eps} by \Ref{Shifman}.  We input 
$F_\pi =93~{\rm MeV}$, $M_{\rho}=770~{\rm MeV}$ and the mass of the lowest-lying vector excited
state, $\rho' (1450)$: $1465\pm 25~{\rm MeV}$\ref{pdg}. The solution to
\Eq{opejunk} with these inputs is given in Table 1. We have included
predictions for the first few vector and axial-vector states in the towers of
states. Of course the entire spectrum is predicted once $M$ and $\Lambda$ are
determined. We have also included a prediction for the chiral perturbation
theory parameter $L_{10}$, which is determined by the mixing angle
$\phi$\ref{Beane2}.  The error quoted in the predicted quantities is from
$\rho' (1450)$. There are, of course, many additional sources of error,
including $1/\nc$ corrections, which conservatively generate an error of 
$\pm 3{F_\pi}$~(the delta-nucleon mass splitting).  The predicted values
of the quark and gluon condensates are somewhat large compared to the 
independent determinations of \Ref{Stern} and \Ref{Indu}, respectively.

\table{gl}
\ninepoint
\caption{\tenpoint
Predictions compared to data using $F_\pi =93~{\rm MeV}$, $M_{\rho}=770~{\rm
MeV}$ and $M_{\rho'}=1465\pm 25~{\rm MeV}$ as input \ref{pdg}. The theoretical
errors are from $M_{\rho'}$. All masses and decay constants are in ${\rm MeV}$,
$L_{10}$ is dimensionless, $\phi$ is in degrees, the gluon condensate is in units of
${\rm GeV}^4$ and the quark condensate (squared) is in units of
$10^{-4}~{\rm GeV}^6$.}
\doublespaced
\ruledtable
                    | TH   | EXP  \dbl                          | TH  | EXP   \cr
$M_{\rho'}$ | INPUT | $1465\pm 25$ \dbl $\phi$ | $44.1\pm 0.6$ | $37.4\pm1.2$\cr
$M_{a_1'}$ | $1465\pm 25$ | $1640\pm 12\pm 30$ \dbl $F_{a_{1}}$ | $96\pm 2$ | $122\pm 23$\cr
$M_{\rho''}$ | $1903\pm 45$ | $1700\pm 20$ \dbl $M_{a_1}$ | $1073\pm 7$ |  $1230\pm 40$\cr
$M_{a_1''}$ | $1903\pm 45$ | NO DATA \dbl $F$ | $137\pm5$ | NO DATA \cr
$M_{\rho'''}$ | $2258\pm 56$ | $2149\pm 17$ \dbl ${\pi{\alpha_s}}\vev{{\bar q}q}^2$ | 
$14.8\pm 0.02$ |  $9\pm 2 $~\ref{Stern}\cr
$M_{a_1'''}$ | $2258\pm 56$ | NO DATA \dbl
${{\alpha_s}}\vev{G_{\mu\nu}G_{\mu\nu}}$ | $0.063\pm 0.006$ | $0.048\pm
0.03$~\ref{Indu}
 \cr
$L_{10}\,(10^{-3})$ | $-5.5\pm 0.1$ | $-5.5\pm 0.7$ \dbl
$\Lambda$ | 
$1172\pm 43~{\rm MeV}$  | $--$
\endruledtable
\endtable


\vskip0.1in
\noindent {\twelvepoint{\bf 4.\quad Scale Setting and Universality}}
\vskip0.1in

\noindent A crucial issue of principle and practice is that of setting the
scale at which the strong coupling constant and Wilson coefficients are
evaluated. Intuitively one would expect that this scale, which can be taken to
represent the onset of perturbative QCD, is a universal parameter, common to
correlators with any given set of quantum numbers. Proving that such a
universal scale exists is highly non-trivial and we do not purport to do so
here. Nevertheless, we will present a plausibility argument that such a
universal parameter is present in large-$\nc$ QCD.  Consider the basic relation
between the hadronic description and the perturbative logarithm in large-$\nc$
QCD. From \Eq{full} and \Eq{cacauno}, neglecting the counterterms,  we have

$$
\, {\sum _{n=0}^{N_{i}}}
{{{{F_{i}^2}}\over{{Q^2}+{M_{i}^2}+n{\Lambda_{i}^2}}}
\,\,\,\mapright{{Q^2}\rightarrow\infty}\, \,\,
-{{F_{i}^2}\over{\Lambda_{i}^2}}}\log{{Q^2}\over{N_{i}{\Lambda_{i}^2}}} \,+\, \ldots.
\EQN scaleset
$$
Here the subscript $i$ is a collective meson quantum number and the dots
represent terms that are suppressed by inverse powers of $Q^2$ and
$N_i$. The perturbative logarithm does not depend on the hadronic
scale $M_i$.  The
natural scale at which to evaluate $\alpha_s$, which we interpret
as the onset of perturbation theory, is $\sim\Lambda_i$.  Naively, the
$\Lambda_i$ can be widely disparate. This renders choosing the physical scale
at which to evaluate $\alpha_s$ channel dependent.
Now consider the following argument.  Regulating QCD with an ultraviolet
cutoff $\Lambda_\infty$ implies

$$
\Lambda_\infty^2=N_{\sss 1}{\Lambda_{\sss 1}^2}=N_{\sss 2}{\Lambda_{\sss 2}^2}=
N_{\sss 3}{\Lambda_{\sss 3}^2}=N_{\sss 4}{\Lambda_{\sss 4}^2}=\ldots \, .
\EQN scaleset2
$$
This is simply the statement that in perturbative QCD there is one scale.
While physics is independent of the cutoff $\Lambda_\infty$, seemingly physics
depends on the $N_i$.  Say we choose to regulate the sums with $N_{\sss
1}=N_{\sss 2}$. \Eq{scaleset2} then implies a new relation between physical
quantities, $\Lambda_{\sss 1}=\Lambda_{\sss 2}$, which is not present if
instead we choose $N_{\sss 1}\neq N_{\sss 2}$.  This ambiguity is removed if

\figure{hybrid.eps}
\epsfxsize 6.5in
\centerline{\epsfbox{hybrid.eps}}
\caption{\tenpoint{The statement of duality for the vector and
axial-vector hybrid correlators. The symbols are explained in \Fig{vanda.eps}.}}
\endcaption
\endfigure

$$
{\Lambda_{\sss 1}}={\Lambda_{\sss 2}}=
{\Lambda_{\sss 3}}={\Lambda_{\sss 4}}=\ldots \equiv\Lambda \, .
\EQN scaleset3
$$
There is then a universal scale, $\Lambda$, which is shared by all large-$\nc$ mesons.
This common physical scale marks the onset of perturbation theory. 
In what follows we will treat \Eq{scaleset3} as a conjecture. Using the
two-loop beta function in the large-$\nc$ limit in $\overline{MS}$ with
$\Lambda{\sss \overline{MS}}=380~{\rm MeV}$, we find 
$\alpha_s (\Lambda )=0.11$, where we have taken $\Lambda =1172\pm 43~{\rm MeV}$ from the analysis
of Section 3 (see Table 1). We will use this value of $\alpha_s$ in the next
section.  One may wonder whether this scale is sensible. If we assume an
underlying string description, this scale relates to the string tension via 
$\Lambda =\sqrt{2\pi\sigma}$. From our analysis in the previous section, taking
$\Lambda$ from Table 1, we then infer $\sqrt{\sigma}=468\pm 12~{\rm MeV}$.
This is in agreement with lattice QCD determinations of the string
tension\ref{Teper}. Lattice calculations of $M_\rho/\sqrt{\sigma}$ from the GF11~\ref{GF11}
and UKQCD~\ref{UKQCD} collaborations yield
$\sqrt{\sigma}=440\pm 15\pm 35~{\rm MeV}$, where the first error is statistical
and the second is systematic\ref{Teper}.
Of course the universality described here is precisely what one expects from Regge theory
arguments, and, indeed, the observed Regge slopes imply
$\sqrt{\sigma}=400-450~{\rm MeV}$~\ref{Teper}.

\vskip0.1in
\noindent {\twelvepoint{\bf 5.\quad Vector and Axial-Vector Currents with Glue}}
\vskip0.1in

\noindent Consider the vector and axial-vector currents

\offparens
$$\EQNalign{ 
&{{\cal V}_{\mu}^a}=
{g_s}{\bar
{q}}{\lambda^\alpha}{G^\alpha_{\mu\nu}}{\gamma^\nu}{\textstyle{1\over
2}}{\tau^a}{q}\, , \EQN hybsdefinqcd;a \cr
&{{\cal A}_{\mu}^a}=
{g_s}{\bar
{q}}{\lambda^\alpha}{G^\alpha_{\mu\nu}}{\gamma^\nu}{\gamma_5}{\textstyle{1\over
2}}{\tau^a}{q}, \EQN hybsdefinqcd;b \cr}$$ 
where $g_s$ is the QCD coupling constant, $G_{\mu\nu}$ is the gluon
field-strength tensor and ${\lambda^\alpha}$ is an $SU(\nc )$ generator.
These currents transform as $(\bf{3},\bf{1})\oplus (\bf{1},\bf{3})$ with
respect to $SU({2})\times SU({2})$.
Since these currents are not conserved, ${{\cal V}_{\mu}^a}$ has nonvanishing
matrix elements between $1^{-+}$ and $0^{++}$ states, and ${{\cal A}_{\mu}^a}$
has nonvanishing matrix elements between $1^{+-}$ and $0^{--}$ states. At
leading order in the $1/\nc$ expansion, the 
$0^{++}$ and $1^{+-}$ states also couple to currents with $n_g=0$.
Therefore we will have
nothing to say about these states here. There are no obvious chiral
constraints between the $1^{-+}$ and $0^{--}$ hybrid states. The nonvanishing 
matrix elements between the hybrid states and the vacuum are

\table{gl}
\tenpoint
\caption{Predictions of $1^{-+}$  hybrids using experimental candidates as
input. On the left (right), we assume that the $\sim 1.4~{\rm GeV}$\ref{exp1} 
~($\sim 1.6~{\rm GeV}$\ref{exp2}) candidate is the ground state.
Theoretical numbers are from \Eq{opejunkhybs} using (A) predicted 
condensates from Table 1 and (B) quark and gluon condensates from \Ref{Stern} and
\Ref{Indu}, respectively (see Table 1). } 
\doublespaced
\ruledtable
                    | TH (A,B)   | EXP  \dbl                           TH (A,B)  | EXP   \cr
$M_{\sss {\hat V}}$ | INPUT | $1392\pm 25$\ref{exp1}  \dbl  INPUT | $1593\pm 8$\ref{exp2} \cr
$\Lambda_{\sss {\hat V}}$ | $1421,1543$ | $--$ \dbl  $1787,1884$ | $--$\cr
$M_{\sss {\hat V}'}$ | $1990,2078$ | $1593\pm 8$\ref{exp2}  \dbl  $2394,2467$ |  NO DATA\cr
$M_{\sss {\hat V}''}$ | $2445,2588$ | NO DATA \dbl $2987,3104$ | NO DATA 
\endruledtable
\endtable

\offparens
$$\EQNalign{ 
&\bra{0}{{\cal V}_{\mu}^a}\ket{{\hat V}^{b}}^{\sss (\lambda )}=
{\delta^{ab}}\,{F_{\sss {\hat V}}}{M^3_{\sss {\hat V}}}\,{\epsilon_\mu^{\sss
(\lambda )}} \, ,\EQN aandvdef;a \cr
&\bra{0}{{\cal A}_{\mu}^a}\ket{{\hat P}^{b}}=
{\delta^{ab}}\,{F_{\sss {\hat P}}}{M^2_{\sss {\hat P}}}\,{p_\mu}.
\EQN aandvdef;b\cr}
$$ 
Here $\hat V$ denotes a $1^{-+}$ hybrid and $\hat P$ denotes a $0^{--}$ hybrid.
The $0^{++}$ and $1^{+-}$ states are denoted $\hat S$ and $\hat A$, respectively.
The relevant spectral functions are defined through

\figure{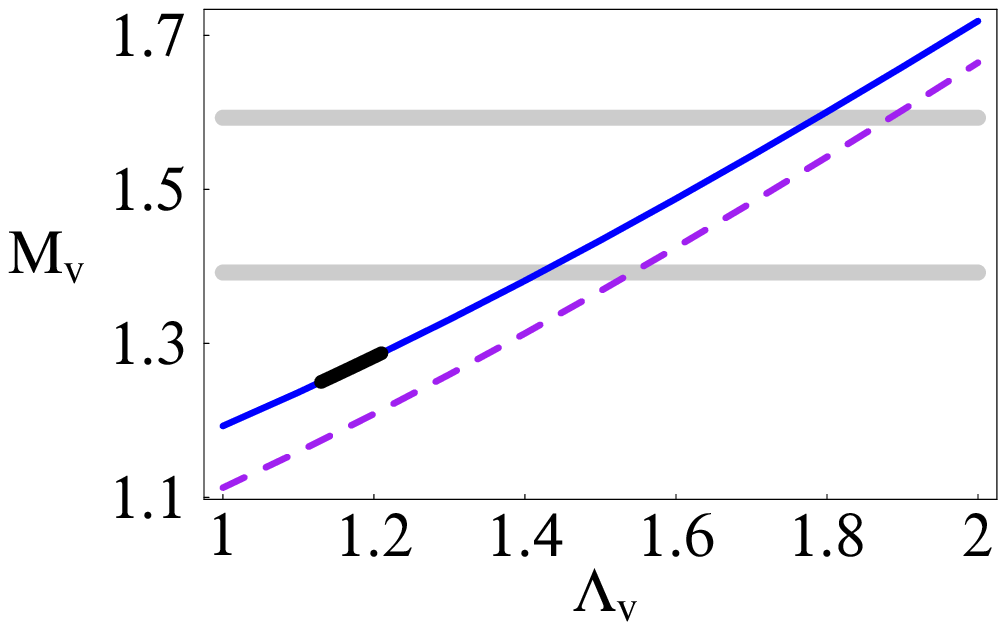}
\centerline{\epsfxsize 3.2in \epsfbox{garbage.eps}\epsfxsize 3.2in \epsfbox{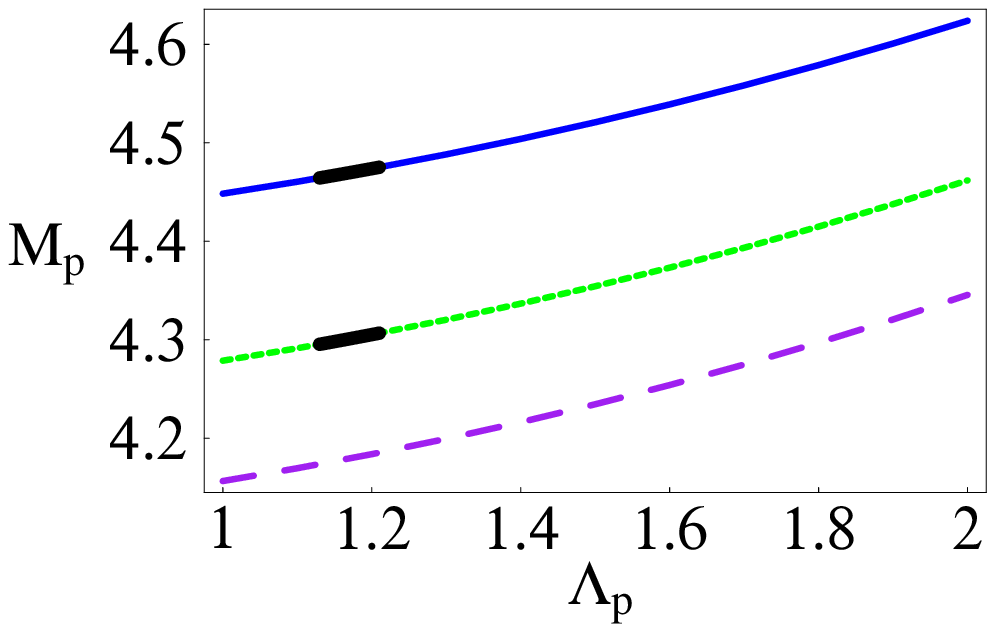}}
\caption{\tenpoint{The left (right) panel gives $M_{\sss {\hat V}}$ ($M_{\sss {\hat P}}$)
as a function of $\Lambda_{\sss {\hat V}}$ ($\Lambda_{\sss {\hat P}}$)
as given by \Eq{opejunkhybs}. The shaded horizontal bars give the experimental
$1^{-+}$ events at $\sim 1.4~{\rm GeV}$\ref{exp1} and $\sim 1.6~{\rm GeV}$\ref{exp2}. 
The solid curves use the predicted condensates from Table 1
and the dashed curves use the quark and gluon condensates from \Ref{Stern} and
\Ref{Indu}, respectively (see Table 1). The blocks over the solid curves are
the universality predictions where $\Lambda$ is taken from Table 1. They
correspond to $\sqrt{\sigma}=468\pm 12~{\rm MeV}$. The dotted curve
in the right panel is evaluated with $\alpha_s =0.13$.}}
\endcaption
\endfigure

$$\EQNalign{ 
&8i\int {d^4x}{e^{iqx}}\bra{0}T\lbrack{{\cal V}_{a}^\mu}(x){{\cal V}_{b}^\nu}(0)\rbrack\ket{0} 
={\delta_{ab}}({q^\mu}{q^\nu}-{g^{\mu\nu}}{q^2}){\Pi_{\sss {\hat V}}}({Q^2}) 
+{\delta_{ab}}{q^\mu}{q^\nu}{\Pi_{\sss {\hat S}}}({Q^2})\, , \EQN correlhyb1;a\cr
&8i\int {d^4x}{e^{iqx}}\bra{0}T\lbrack{{\cal A}_{a}^\mu}(x){{\cal A}_{b}^\nu}(0)\rbrack\ket{0} 
={\delta_{ab}}({q^\mu}{q^\nu}-{g^{\mu\nu}}{q^2}){\Pi_{\sss {\hat A}}}({Q^2}) 
+{\delta_{ab}}{q^\mu}{q^\nu}{\Pi_{\sss {\hat P}}}({Q^2}). \EQN correlhyb1;b\cr}
$$ 
Since $\Pi_{\sss {\cal V},{\cal P}}({Q^2})$ are
dimension four, we are here interested in the case $m=2$. The spectral
functions in the large-$\nc$ limit are 

$$\EQNalign{ 
&{1\over\pi} Im \,\Pi_{\sss{\hat V}}(t)= 
2\sum_{n=0}^{N_{\sss {\hat V}}} {F_{\sss {\hat V}}^2}(n){M_{\sss{\hat V}}^4}(n)
\delta (t-M_{\sss {\hat V}}^2 (n)) \, ,
\EQN spectfunhybs;a \cr
&{1\over\pi} Im \,\Pi_{\sss{\hat P}}(t)= 
2\sum_{n=0}^{N_{\sss {\hat P}}} {F_{\sss {\hat P}}^2}(n){M_{\sss{\hat P}}^4}(n)
\delta (t-M_{\sss {\hat P}}^2 (n)) .
\EQN spectfunhybs;b \cr}
$$

Reading directly from \Eq{match1;a} and \Eq{matchm2} we find the system of equations

\autoparens
$$\EQNalign{ 
&\qquad\qquad{\cal C}_{\sss {\hat V}}=-{{\alpha_s}\over{60{\pi^3}}}\, ,\qquad\qquad
{\cal W}_{\sss {\hat V}}=-{{\alpha_s}\over{9\pi}}\vev{G_{\mu\nu}G_{\mu\nu}}\, ,
\EQN opejunkhybs;a\cr
&\qquad\qquad{\cal C}_{\sss {\hat P}}=-{{\alpha_s}\over{120{\pi^3}}}\, ,\qquad\qquad
{\cal W}_{\sss {\hat P}}={{\alpha_s}\over{6\pi}}\vev{G_{\mu\nu}G_{\mu\nu}}\, ,
\EQN opejunkhybs;b\cr
&{\vev{{\cal O}}^{\sss d=6}_{\sss {\hat V}}}=
-{4\over 9}{\pi{\alpha_s}}\vev{{\bar q} q}^2=
({1\over 3}{\cal C}_{\sss {\hat V}}({M_{\sss {\hat V}}^2}-
{\Lambda_{\sss {\hat V}}^2}){M_{\sss {\hat V}}^2}
+{\cal W}_{\sss {\hat V}})
( {M_{\sss {\hat V}}^2}-{1\over 2}{\Lambda_{\sss {\hat V}}^2})\, ,
\EQN opejunkhybs;c\cr
&{\vev{{\cal O}}^{\sss d=6}_{\sss {\hat P}}}=
-{4\over 3}{\pi{\alpha_s}}\vev{{\bar q} q}^2=
({1\over 3}{\cal C}_{\sss {\hat P}}({M_{\sss {\hat P}}^2}-
{\Lambda_{\sss {\hat P}}^2}){M_{\sss {\hat P}}^2}+
{\cal W}_{\sss {\hat P}})
({M_{\sss {\hat P}}^2}-{1\over 2}{\Lambda_{\sss {\hat P}}^2})\, .
\EQN opejunkhybs;d\cr}
$$ 
The Wilson coefficients and ${\cal C}_{\sss {{\hat V},{\hat P}}}$ have been computed
from the diagrams in \Fig{hybrid.eps} by \Ref{Govaerts}. The masses of the
lowest-lying hybrids, ${M_{\sss {\hat V},{\hat P}}}$, require knowledge
of ${\Lambda_{\sss {\hat V},{\hat P}}}$. The purest predictions we can make use
the $1^{-+}$ experimental candidates as input to determine ${\Lambda_{\sss {\hat
V}}}$. These results are given in Table 2. We give predictions using the
condensates extracted in the analysis of Section 3, and from independent
determinations\ref{Stern}\ref{Indu}~(See Table 1). The spread
in values gives a 
sense of the theoretical error due to truncating the OPE.
In \Fig{garbage.eps} we plot
$M_{\sss {\hat V}}$ ($M_{\sss {\hat P}}$) as a function of 
$\Lambda_{\sss {\hat V}}$ ($\Lambda_{\sss {\hat P}}$) as given by \Eq{opejunkhybs}.
We give curves for both sets of condensate parameters. The sensitivity to
$\alpha_s$ is insignificant for $M_{\sss {\hat V}}$. In \Fig{garbage.eps} we
show a curve for $M_{\sss {\hat P}}$ evaluated with $\alpha_s=0.13$.

\table{gl}
\tenpoint
\caption{Predictions for $1^{-+}$  and $0^{--}$ hybrids using universality. As
input we use 
$\Lambda_{\sss {\hat V}}=\Lambda_{\sss {\hat P}}=\Lambda=1172\pm 43~{\rm
MeV}$~(see Table 1), which corresponds to a string tension of 
$\sqrt{\sigma}=468\pm 12~{\rm MeV}$.} 
\doublespaced
\ruledtable
                    | TH    | EXP  \dbl     |       TH   | EXP   \cr
$M_{\sss {\hat V}}$ | $1190\pm 20$ | $1392\pm 25$ \dbl $M_{\sss {\hat P}}$ |
$4180 \pm 5$ | NO DATA\cr
$\Lambda_{\sss {\hat V}}$ | INPUT | $--$ \dbl  $\Lambda_{\sss {\hat P}}$
| INPUT | $--$\cr
$M_{\sss {\hat V}'}$ | $1670\pm 40$ | $1593\pm 8$ \dbl  $M_{\sss {\hat P}'}$|
$4340\pm 16$ |  NO DATA\cr
$M_{\sss {\hat V}''}$ | $2050\pm 50$ | NO DATA \dbl $M_{\sss {\hat P}''}$
|$4495\pm 26$ | NO DATA 
\endruledtable
\endtable

\vskip0.1in
\noindent {\twelvepoint{\bf 6.\quad Discussion}}
\vskip0.1in

\noindent If we treat the $1^{-+}$ experimental candidate at $\sim 1.4~{\rm GeV}$\ref{exp1} as
the lowest-lying state, then we predict the first excited state at 
$\sim 2~{\rm GeV}$, as compared to $\sim 1.6~{\rm GeV}$\ref{exp2}~(see Table 2). This
discrepancy is larger than one would expect from a $1/\nc$ correction.
If we treat the $1^{-+}$
experimental candidate at $\sim 1.6~{\rm GeV}$\ref{exp2} as the lowest-lying state, then we
predict the first excited state at $\sim 2.4~{\rm GeV}$. One must then
understand the role of the $\sim 1.4~{\rm GeV}$\ref{exp1} experimental candidate. We reiterate that in
the large-$\nc$ limit, mesons do not couple to QCD currents with four
quarks. If such mixing is relevant, it corresponds to a large subleading effect
in the $1/\nc$ expansion.
Without assuming universality, there is no prediction for the lowest-lying
hybrid states. With $\Lambda_{\sss {\hat V}}$ and $\Lambda_{\sss {\hat P}}$
between $1-2~{\rm GeV}$, $M_{\sss {\hat V}}$ ranges between 
$1.1-1.7~{\rm GeV}$, while $M_{\sss {\hat P}}$ ranges between $4.1-4.7~{\rm GeV}$~(See
\Fig{garbage.eps}). 
Assuming universality, and taking
$\Lambda_{\sss {\hat V}}=\Lambda_{\sss {\hat P}}=\Lambda=1172\pm 43~{\rm MeV}$
from the analysis of Section 3, we find a lowest-lying hybrid 
$\sim 1.2~{\rm GeV}$, which underestimates the mass of the low-lying
experimental candidate, and a first excited state
$\sim 1.7~{\rm GeV}$, which overestimates the mass of the heavier experimental
candidate~(see Table 3). However, the discrepancies are consistent with expected $1/\nc$ 
corrections.


\vskip0.1in
\noindent {\twelvepoint{\bf Acknowledgments}}
\vskip0.1in

\noindent I thank Tom Cohen, Maarten Golterman, Santi Peris, Martin Savage and
Eric Swanson for valuable discussions. I am grateful to Nathan Isgur for
suggesting this project.  This work was supported by the U.S. Department of
Energy grant DE-FG03-97ER-41014.

\nosechead{References}
\ListReferences \vfill\supereject \end